\documentclass[useAMS,usenatbib,usegraphicx]{mn2e}

\usepackage{psfig}
\usepackage{times}
\usepackage{subfigure}

\newcommand{\kms}{\hbox{km s$^{-1}$}}

\newcommand{\vsini}{\hbox{$v$\,sin\,$i$}}

\newcommand{\degs}{$\degr$}
\newcommand{\chisq}{$\chi^{2}$}

\newcommand{\radday}{\hbox{rad.day$^{-1}$}}

\newcommand{\ha}{H$\alpha$}

\newcommand{\speedy}{\hbox{Speedy Mic}}

\begin{document}

\title[Speedy Mic prominence system - I]{The coronal structure of Speedy Mic - I: A densely packed prominence system beyond co-rotation}

\makeatletter
 
\def\newauthor{%
  \end{author@tabular}\par
  \begin{author@tabular}[t]{@{}l@{}}}
\makeatother

\author[N.J.~Dunstone, J.R.~Barnes, A.~Collier~Cameron, M.~Jardine]
{N.J.~Dunstone$^1$\thanks{E-mail: njd2@st-andrews.ac.uk} J.R. Barnes$^1$ A. Collier Cameron$^1$ M. Jardine$^1$\\
$^1$ School of Physics and Astronomy, University of St Andrews, Fife KY16 9SS. UK. }

\date{2005, 200?}

\maketitle

\begin{abstract}
We present new observations of the prominence system on the K3 dwarf \speedy\ (BO Mic, HD 197890).  Using an improved technique to track the absorption features in \ha\ we find a very active prominence system with approximately ten prominences on the observable hemisphere per rotation.  From a total of 25 prominences we find an average axial distance of $(2.85\pm0.54) R_*$ which is
twice the co-rotation radius above the stellar surface.  We discuss the consequences of these observations on the nature of the supporting magnetic structures.  Two consecutive nights, with complete phase coverage, combined with a further night after a three night gap allow us to study the evolution of the prominence system on two different timescales.  { {Several of the prominences have}} counterparts at similar phases on consecutive nights.  During this interval many prominences show evidence for evolution in their heights and phases of observation.  Five nights (13 rotation cycles) later we recover many prominences at approximately the same phases.  Whilst individual prominences change axial distances or appear/re-appear from night-to-night the underlying prominence supporting structures appear to be stable over as many as 13 stellar rotations.  
\end{abstract}


\section{INTRODUCTION}
Studies of young rapidly rotating cool stars have revealed many solar analogues.  Advances in Doppler imaging techniques \citep{vogt83spot} have now shown a number of highly spotted stars indicating efficient magnetic dynamo activity \citep{strassmeier02survey}. More recently Zeeman-Doppler Imaging (ZDI) has provided the first magnetic surface maps of these active stars \citep{donati97ab}.  In order to fully understand the field topology of rapidly rotating stars, we need to examine the relationship between surface and coronal magnetic fields.  Potential field extrapolations have made use of magnetic images to model the coronal magnetic structure \citep{jardine02xray}. This enables a comparison to be made with the hot extended active stellar coronae inferred from X-ray observations.  Here we present observations of stellar prominences which give us a direct probe into the coronal magnetic structure and compare our results in the context of field configurations.

Stellar prominences are observed as rapid spectroscopic variations in the stellar Balmer lines.  Transient absorption features passing through the Doppler broadened \ha\ line of rapidly rotating stars were first reported by \cite{cam1989a} on the rapidly rotating K0V dwarf AB Doradus (AB Dor).  These were found to be cool dense condensations of material in enforced co-rotation with the stellar surface and so dubbed ``prominences".  

From well sampled \ha\ timeseries one can infer the spatial distribution of the prominence structures from the rate at which absorption features are seen to drift through the profile.  Such observations have been carried out by different authors on around a dozen stars over the last two decades.  For a recent review of prominence observations see \cite{cam2003rev}.  AB Dor however, remains the most studied object to date. Prominences have consistently been reported (\citealt{cam1989a}; \citealt{cam1989b}, \citealt{donati97ab}; \citealt{cam1999}; \citealt{donati99ab}) to lie at a range of distances from the stellar rotation axis from 2 to 8 $R_*$, with a concentration at the co-rotation radius (2.7$R_*$ on AB Dor). 

As their solar namesake suggests these stellar prominences are confined and supported by magnetic loop structures.  This is an immediate result of finding prominences beyond the co-rotation radius where a magnetic tension term is needed to balance the outward centrifugal force.  Therefore observations of prominence locations and evolution provide a unique probe to the stellar magnetic field out to distances of many stellar radii into the corona.

Here we present new observations of the prominence system on one of the fastest and brightest young rapidly rotating stars in the solar neighbourhood.  Speedy Mic (BO Mic, HD 197890) is a K3V dwarf with an equatorial rotation velocity \hbox{\vsini\ $\simeq$ 132 \kms} \citep{barnes05}.  Recent studies of X-ray variability have shown Speedy Mic to be one of the most active stars in the solar neighbourhood. \cite{singh99xray} found $\rm{log}({L_X}/{L_{bol}})\simeq-3$ placing it in the saturated regime. More recently \cite{marakov03xray} reported $\rm{log}({L_X}/{L_{bol}})\simeq-2$ although this may be due to flaring as otherwise it would put Speedy Mic's X-ray luminosity an order of magnitude greater than any other star in its class.

Prominences were first observed on Speedy Mic by \cite{jeffries1993}. Two prominences were reported at radial distances of $1.36\pm0.09R_*$ and $2.96\pm0.14 R_*$.  \cite{barnes2001} also reported four prominences at radial distances in the range $1.3<R<3.8R_*$. 

In \S \ref{sect:obs} we present observations with uninterrupted and complete phase coverage of the Speedy Mic \ha\ line on consecutive nights.  This strategy is combined with an improved technique for tracking prominences described in \S \ref{sect:promtr}.  The relatively large number of resulting prominences allow us to examine the distribution of prominence heights and evolution timescales in \S \ref{sect:proman}.  We discuss the implications these results have on the nature of supporting magnetic structures in \S \ref{sect:disc}.  Finally we present our conclusions in \S \ref{sect:conc}.

\section{OBSERVATIONS}
\protect\label{sect:obs}

Observations were made with the University College London \'{E}chelle Spectrograph (UCLES) at the Anglo Australian Telescope (AAT) on 2002 July 18, 19 \& 23. { {A slit width of \mbox{1.2 \arcsec}~ was used, providing a mean resolution of $\sim$ 45000 at \ha\ with coverage from 4376\AA~to 6892\AA~in 47 orders on the 2Kx4K EEV CCD.}}

{ {Full phase coverage was obtained on both July 18 and 19 corresponding to 100 and 97 spectra respectively. The typical signal-to-noise (S/N) achieved was 80 in a 300s exposure.}} On July 23 time was lost due to bad weather and thus only resulted in 67 usable spectra ($\sim70\%$ phase coverage) some of which have poorer S/N.  

\begin{figure}
\begin{center}
\includegraphics[height=8cm,angle=270]{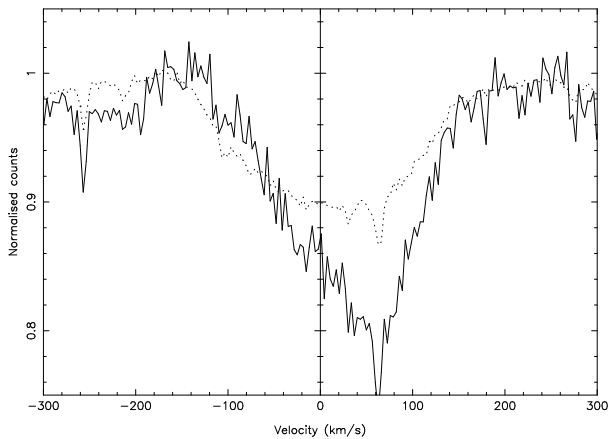}
\end{center}
\caption{An example observed \ha ~profile (solid line) shows an absorption transient. Also plotted is the average profile (dotted line) for comparison.  The sharp telluric line features (e.g. located at 60 \kms) are removed during the unsharp masking process.}
\protect\label{fig:exampha}
\end{figure}

Spectral extraction was performed using ECHOMOP, the \'{e}chelle reduction package developed by \cite{mills92} and its implementation of the optimal extraction algorithm developed by \cite{horne86extopt}.  Further details of the extraction process and a journal of observations can be found in \cite{barnes05}.  Throughout this paper we shall be using the system parameters found using the same dataset by \cite{barnes05}.  These are summarised in Table \ref{tab:sysparam}.  We note that  \cite{wolter05a} find very similar parameters from data taken a few weeks later using the 8-m Very Large Telescope (VLT).

\begin{table}
\caption[System parameters]{System parameters for Speedy Mic from \cite{barnes05}}
\protect\label{tab:sysparam}
\vspace{5mm}
\centering
\begin{tabular}{lc}
\hline                     
P [d]				&  0.38007 $\pm$ 0.00005   \\
$v_{r}$~[kms$^{-1}$]		&  -8.0 $\pm$ 1.0   \\
\vsini 	~[kms$^{-1}$]		&  132 $\pm$ 2 \\
Axial inclination [deg]		&  70.0 $\pm$ 5   \\
\hline
\end{tabular}
\end{table}

We take the \ha\ order and normalise the continuum of each individual spectrum.  Any residual tilt in the spectrum is corrected by using a rigid spline fit after masking out the \ha\ line region.  In Fig. \ref{fig:exampha} an example observed \ha ~profile can be seen with a prominence absorption feature finishing its path across the stellar disc.  The spectra are stacked together and phased according to their times of observation using midnight (UT) of the first night as the epoch of phase zero.

\begin{figure*}
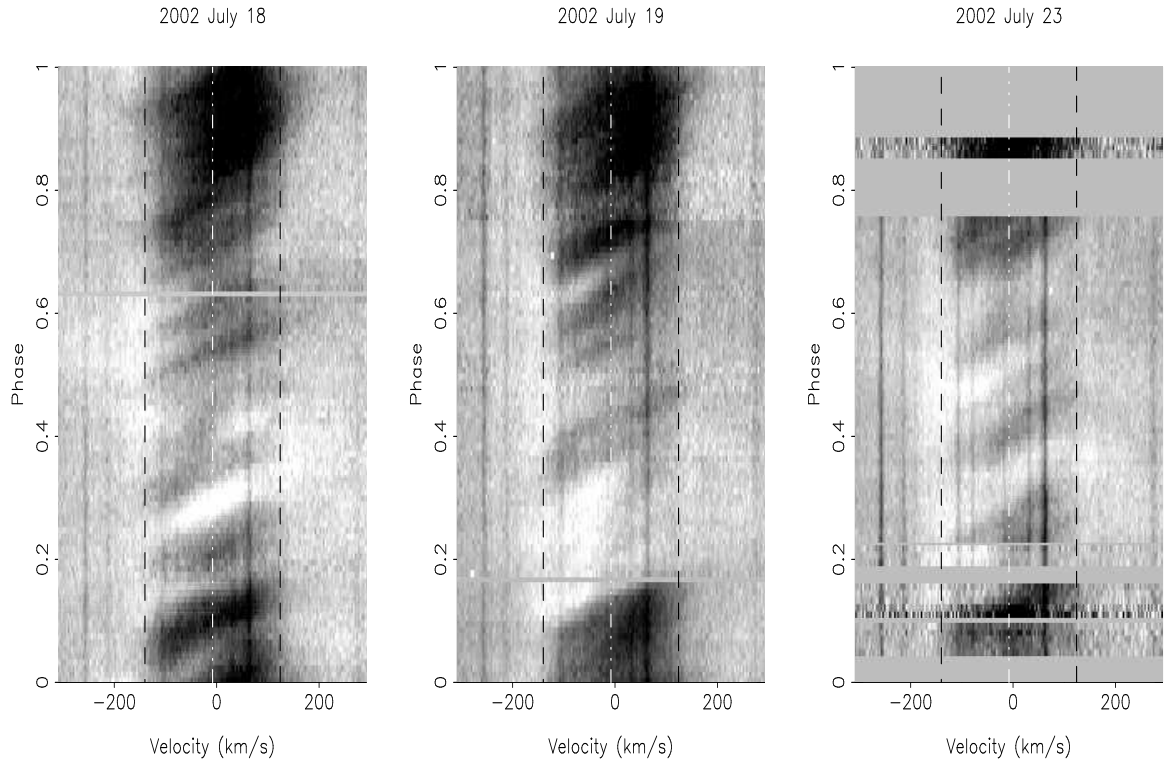

\begin{center}

  \begin{tabular}{ccc}
    \includegraphics[width=5.0cm,height=10cm,angle=0]{Image/njd_speedy_fig2.ps} &
    \hspace{-2mm}
    \includegraphics[width=5.0cm,height=10cm,angle=0]{Image/njd_speedy_fig3.ps} &
     \hspace{-2mm}
    \includegraphics[width=5.0cm,height=10cm,angle=0]{Image/njd_speedy_fig4.ps} \\
  \end{tabular}

\end{center}
\caption[Speedy Mic \ha\ time series spectra ]{Raw \ha\ time series spectra for 2002 July 18, 19 \& 23, with phase plotted against velocity.  Dashed black lines show the \vsini\ limits and the central dashed white line shows the radial velocity of Speedy Mic.  The grey-scale runs from black at 0.80 times the local continuum level, to white at 1.02 times continuum.}
\protect\label{fig:rtseries}
\end{figure*}

The phased raw \ha\ time series are shown in Fig. \ref{fig:rtseries}.  A visual examination of the time series reveals Speedy Mic to have a remarkably active prominence system.  Many rapid transient absorption features are seen to drift through the stellar \ha\ profile.  

The most noticeable feature however is a very dark region between phases $0.8<\phi<1.0$.  Further examination suggests that the timeseries splits into two halves, with a lighter region between  phases $0.2<\phi<0.7$ and the rest of the timeseries being dominated by darker absorption features.  We will briefly discuss these contrasting regions in \S \ref{sect:conc}.  The rest of this paper concerns the rapid transients seen in the \ha\ profile.

\section{PROMINENCE TRACKING}
\protect\label{sect:promtr}

{ {The spatial distribution of prominences can be inferred from the timeseries.  The phase at which any individual prominence transits the center of the stellar rotation profile can be easily identified.  In addition to this by assuming that the prominences are in co-rotation with the surface of \speedy\ (this assumption is validated by re-acquisition of the same prominence on subsequent nights) then we can calculate the axial distances (heights).  This is due to the fact that the drift rates of the absorption transients through the stellar rotation profile are directly proportional to their radial diatnces from the rotation axis. Due to the relatively large heights of the prominences their orbital velocity sinusoids are well approximated by straight lines while they are transiting the stellar disc.

The projected height is given by:

$$\frac{\varpi}{R_*}=\frac{\dot{v}}{\Omega{v}\mathrm{sin}i}$$

where $\varpi=R\mathrm{cos}\theta$ is the distance of a prominence at latitude $\theta$ from the rotation axis. $\dot{v}$ is the observed drift rate through the stellar rotation profile and is obtained directly from the phased timeseries.  $\Omega$ is the stellar angular velocity, $R_*$ is the stellar radius and \vsini\ is the projected equatorial rotation speed.}}

Previous authors have used a number of different methods to estimate the drift rates of the prominence absorption signatures through the line profile.  When the S/N is poor, or there are gaps in the phase coverage of the prominence tracks, estimates can be made directly by drawing a by-eye best fit line through the prominence absorption signatures as was done in \cite{barnes2001}.  However the associated uncertainties were quoted as being $\simeq1R_*$.  With more complete datasets the skew-mapping technique as devised in \cite{cam1989a} and updated by \cite{donati97ab} can be used. In order to bring out just the ridge-lines of the absorption features an unsharp-mask is often performed on the raw \ha\ timeseries.  Further details will be given in the next section.

These techniques do not attempt to use a prior model for the prominences as they pass through the profile.  The advantage of using a model is that we can obtain a confidence assessment through a \chisq\ analysis.  This describes how well the model fits the data and so is not as biased by strong areas of absorption but rather favours a consistent track across the profile.  We describe the use of such a matched filter analysis technique with a Gaussian model in \S \ref{sect:MFA}.  The first stage in the analysis however is to remove the atmospheric telluric lines from the timeseries spectra shown in Fig. \ref{fig:rtseries}, while at the same time bringing out the ridge-lines of the prominences.

\subsection{Unsharp masking}

Previous studies tracking prominences through the \ha\ profile have subtracted or divided by a mean profile in order to remove the telluric lines.  As Fig. \ref{fig:rtseries} shows however the Speedy Mic \ha\ profile is filled with many strong absorption signatures and it is difficult to establish a base reference level for the \ha\ line.  The mean profile of the \ha\ line obtained by collapsing the timeseries in the temporal direction is therefore in strong absorption { {(see Fig. \ref{fig:exampha})}}.  When the timeseries is divided by the mean profile we obtain spurious regions of relative emission, in which weak absorption transients can be seen.  This is undesirable for further analysis as the absorption features should be below the continuum level.  Another problem when using a mean profile is the assumption that the \ha\ profile is symmetric, i.e. every transient absorption feature is symmetric around the centre of the stellar \ha\ profile. Again though just from looking at some of the absorption transients this not the case especially around the $0.8<\phi<1.0$ region where an excess of absorption in the red is seen.  

In order to remove the telluric lines and to bring out the ridge lines of the absorption features we perform an unsharp mask in the temporal direction. We thus smooth the timeseries using a Gaussian-weighted running mean with $\sigma=1500s$.  This corresponds to approximately half the duration of a typical absorption transient.  We then divide the original timeseries by this smoothed version.  The resulting timeseries spectra are shown in Fig. \ref{fig:utseries}.  This process has two main advantages.  Firstly it removes the telluric lines very well as it does so on localised regions rather than the whole timeseries.  This then allows for changes in the strengths of the telluric features as observations are made through varying amounts of the Earth's atmosphere during the night.  Such a procedure also brings out the ridge lines of the transients making them more obvious.  It is important to realise however that this is achieved because this technique essentially simplifies the timeseries, removing the underlying slow variations of the \ha\ profile.  Another point also worth remembering is that the absolute strengths of the absorption (or indeed emission) with respect to the continuum level are not maintained.  However this technique provides us with a higher contrast version of the timeseries which has brought out the rapid absorption transients that we are interested in.

\subsection{Matched filter analysis tracking of prominence features}
\protect\label{sect:MFA}

In order to track the absorption features as they move across the \vsini\ profile of the star we use a matched filter analysis technique.  This is based upon the spot tracking technique of \cite{cam02} adapted for the prominence situation.  We model the absorption features as Gaussians of fixed width moving linearly with time through the \ha\ profile. We adopt  $\sigma=$25\kms which was found to be approximately the width of the weakest absorption transients.  We then sample the timeseries at just over twice the number of spectra, a spacing of ~0.004 in phase { {to ensure precise determination of prominence phases.}}  At each phase the path through the profile of a potential prominence at that meridian crossing is considered for a range of possible prominence heights.  Heights are considered from the stellar surface out to ~7.5$R_\star$ with a resolution of ~0.015$R_\star$ { {to ensure adequate radial resolution for the low-lying prominences located well inside the co-rotation radius}}.  Hence over a hundred thousand possible prominence tracks are tried throughout the timeseries.

At each prominence height the model Gaussian is then optimally scaled to fit the data along the whole prominence track.  The result is a single scale factor for the potential prominence moving through the \vsini\ profile at that particular phase and height, along with an associated \chisq\ for the fit.  

Initially all points that fell between the \vsini\ limits of the star were selected for Gaussian fitting, so that equal weight was given independent of position along the \vsini\ profile.  This resembles a step function along the profile, with a value of one inside the \vsini\ range and zero outside.  This approach was found to be an over-simplification when tested on actual absorption transients.  { {The strength of a given absorption transient is determined by the physical properties of the cloud and its orientation with respect to our line of sight.  If the cloud appears to transit the stellar disc at a high latitude then the extent of the absorption transient in the rotational profile will be reduced.  The finite size of the cloud means that it takes time to move fully onto the disc. This combined with the velocity dispersion within the cloud itself and stellar limb darkening all change the strengths of the absorption features as they cross the disc.} }

Furthermore for prominences at larger heights the path of the absorption signatures samples less spectra.  Thus at the maximum height that we consider, the absorption signature samples only five spectra.  This produces discontinuities in the resulting \chisq map, since as we increase the height we jump from considering {\it{n}} spectra to {\it{n-1}} spectra.

In order to make a more realistic model for the absorption strength across the profile we parameterised all these factors into a single weighting function, shown in Fig. \ref{fig:wfunc}. This took the form of a Fermi-Dirac like function
$$W=\frac{1.0}{\left(1.0+\rm{exp}\left(\alpha\frac{(v_{cent}\pm{v})}{{\Delta}v_{prom}}\right)\right)},$$
where the parameter $\alpha$ controls the nature of the fall-off away from the line centre and ${\Delta}v_{prom}$ is an approximation to the internal prominence velocity dispersion.  We approximated ${\Delta}v_{prom}$ to be roughly twice the width of the fitted absorption features, $\sigma=$25 \kms.  The parameter $\alpha$ was changed to crudely optimise the fit to observed prominences passing through the profile, resulting in the profile shown in Fig. \ref{fig:wfunc}.   This gives less weight to the extremes of the \vsini\ limits and it also produces smooth \chisq\ minima by removing the sharp and discontinuous jumps in the number of spectra sampled.  Note that this function is of course only an approximation as the precise strength of any given feature is unique.

\begin{figure}
\begin{center}
\includegraphics[height=8cm,angle=270]{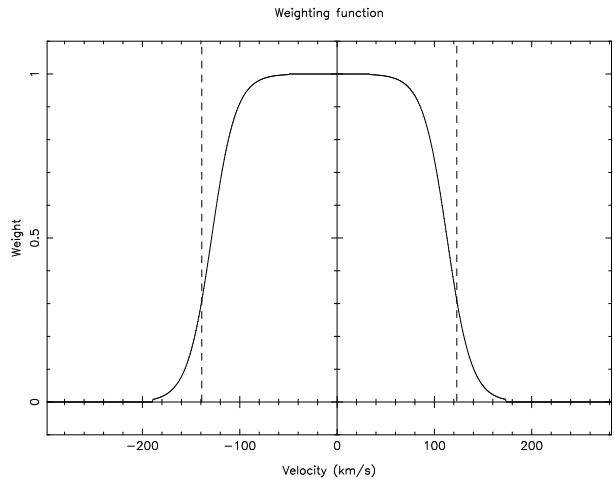}
\end{center}
\caption{The weight given to the scale factor of the Gaussian model as it crosses the stellar \vsini\ profile.  Function is centred on Speedy Mic's radial velocity, $v_r=8$\kms\ }
\protect\label{fig:wfunc}
\end{figure}

Scale factors obtained through this procedure are either negative or positive, corresponding to prominence absorption signatures or emission features respectively.  { {It should be noted}} however that most of the emission features are merely a product of the unsharp-mask process and are therefore not real.  Just relying on the scale factor of the Gaussian is insufficient to reveal the prominence heights.   This is because the scale factor alone will tend to overestimate the height of the prominences as it can join neighbouring prominences together.  { {The heights we find are therefore based upon the cuts (at constant phase) through the resulting \chisq\ map at each phase considered.  The resulting \chisq\ minimum in height is then compared to neighbouring phases (constant height) to establish a local \chisq\ minimum for a particular prominence.}} This results in much better determined minima and are used to produce the prominence tracks that are overlaid on the unsharp masked timeseries in Fig. \ref{fig:utseries}.  Overlaying the tracks upon the timeseries in this manner allows quick visual confirmation that a suitable fit has been achieved.

The uncertainty in the prominence heights can then be estimated from the curvature of the \chisq\ slices.  However due to the asymmetry in the \chisq\ minima the reported uncertainties are mean estimates and almost certainty over-estimate the formal error on the prominence heights.  The results of this improved tracking technique on the Speedy Mic dataset are displayed in Table \ref{tab:results}.

\begin{figure*}
\begin{center}

  \begin{tabular}{ccc}
    \includegraphics[width=5.0cm,height=10cm,angle=0]{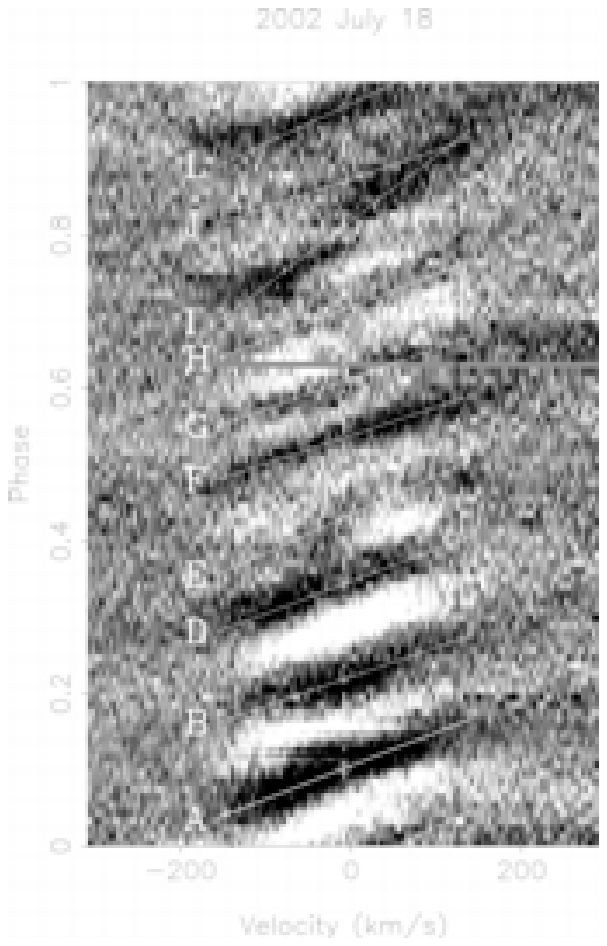} &
    \hspace{-2mm}
    \includegraphics[width=5.0cm,height=10cm,angle=0]{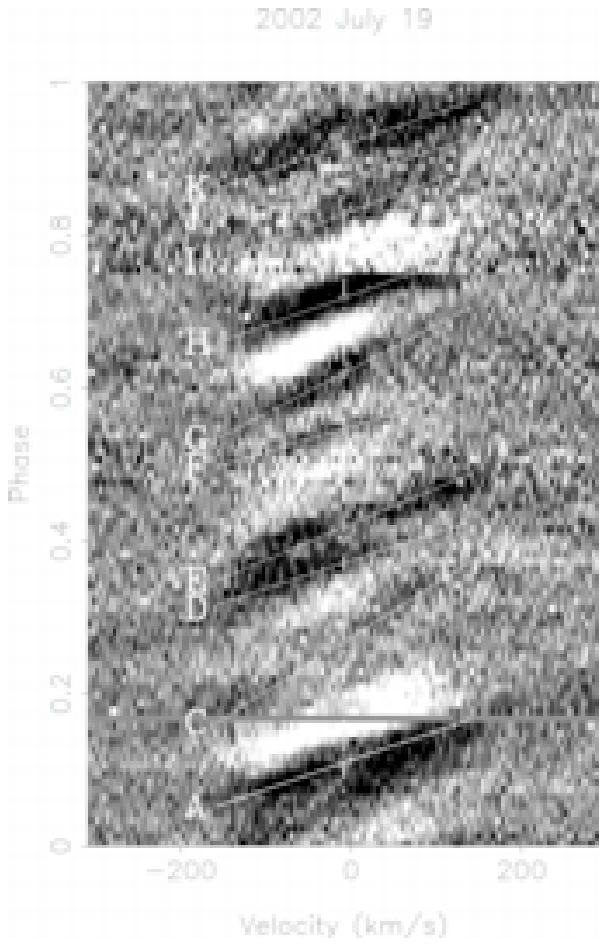} &
     \hspace{-2mm}
    \includegraphics[width=5.0cm,height=10cm,angle=0]{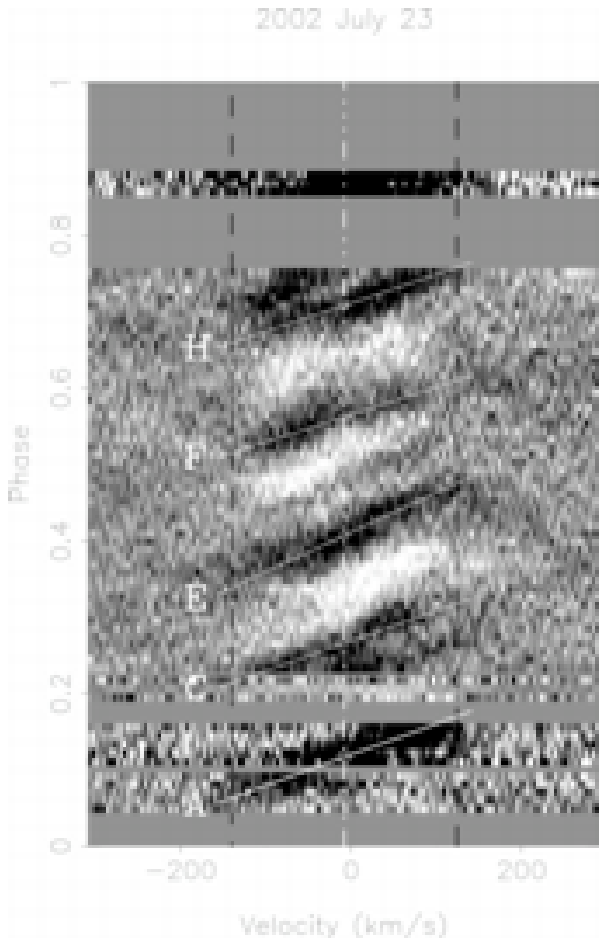} \\
  \end{tabular}

\end{center}
\caption[Speedy Mic \ha\ time series spectra ]{As Fig. \ref{fig:rtseries} but after the unsharp masking process described in the text. The grey-scale now runs from black at 0.97 times the local continuum level to white at 1.03 times continuum. Superimposed are the fits achieved by the prominence tracking technique.}
\protect\label{fig:utseries}
\end{figure*}

\begin{figure*}
\begin{center}

  \begin{tabular}{cc}
    \includegraphics[width=7.0cm,angle=0]{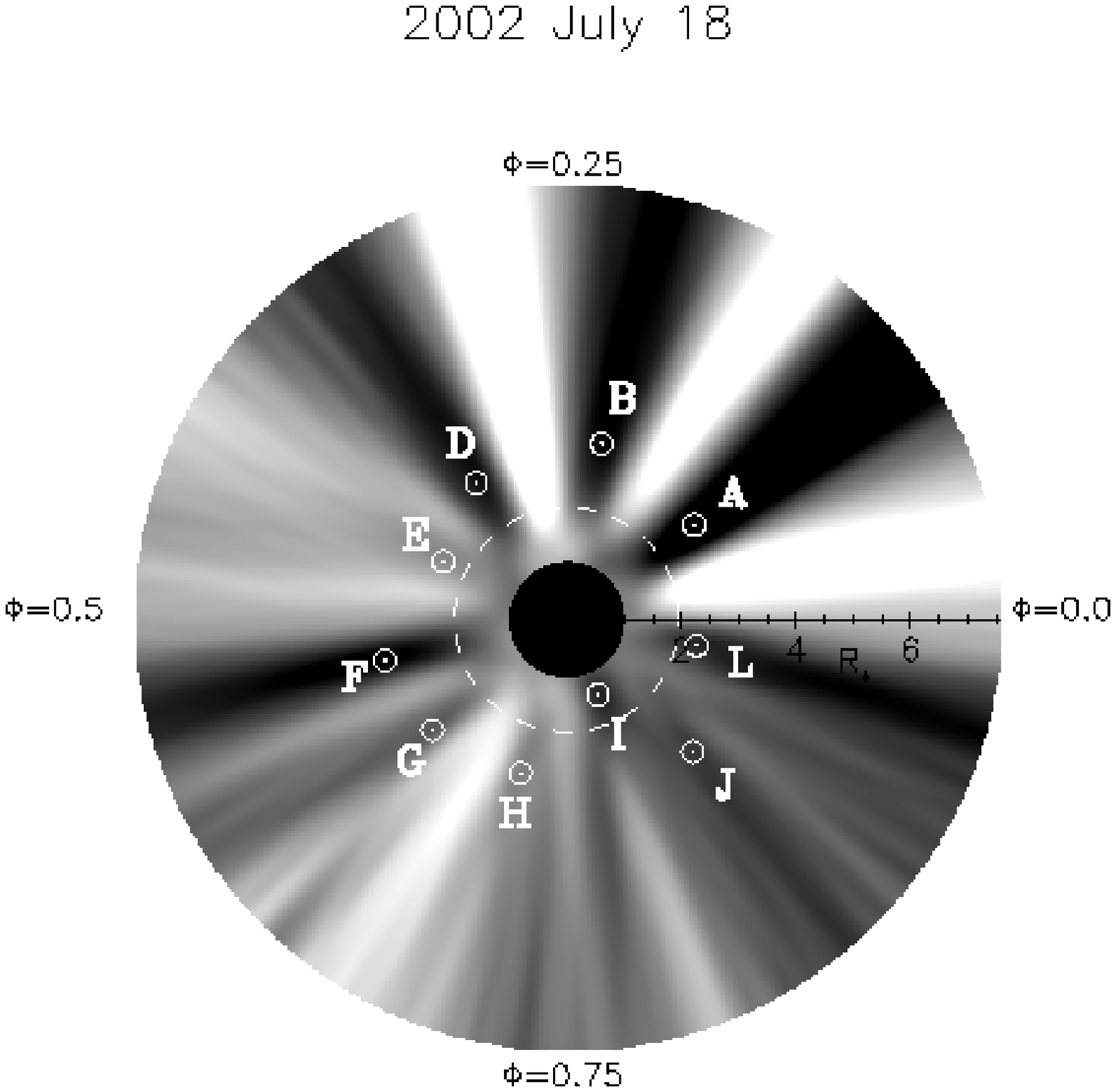} &
    \includegraphics[width=7.0cm,angle=0]{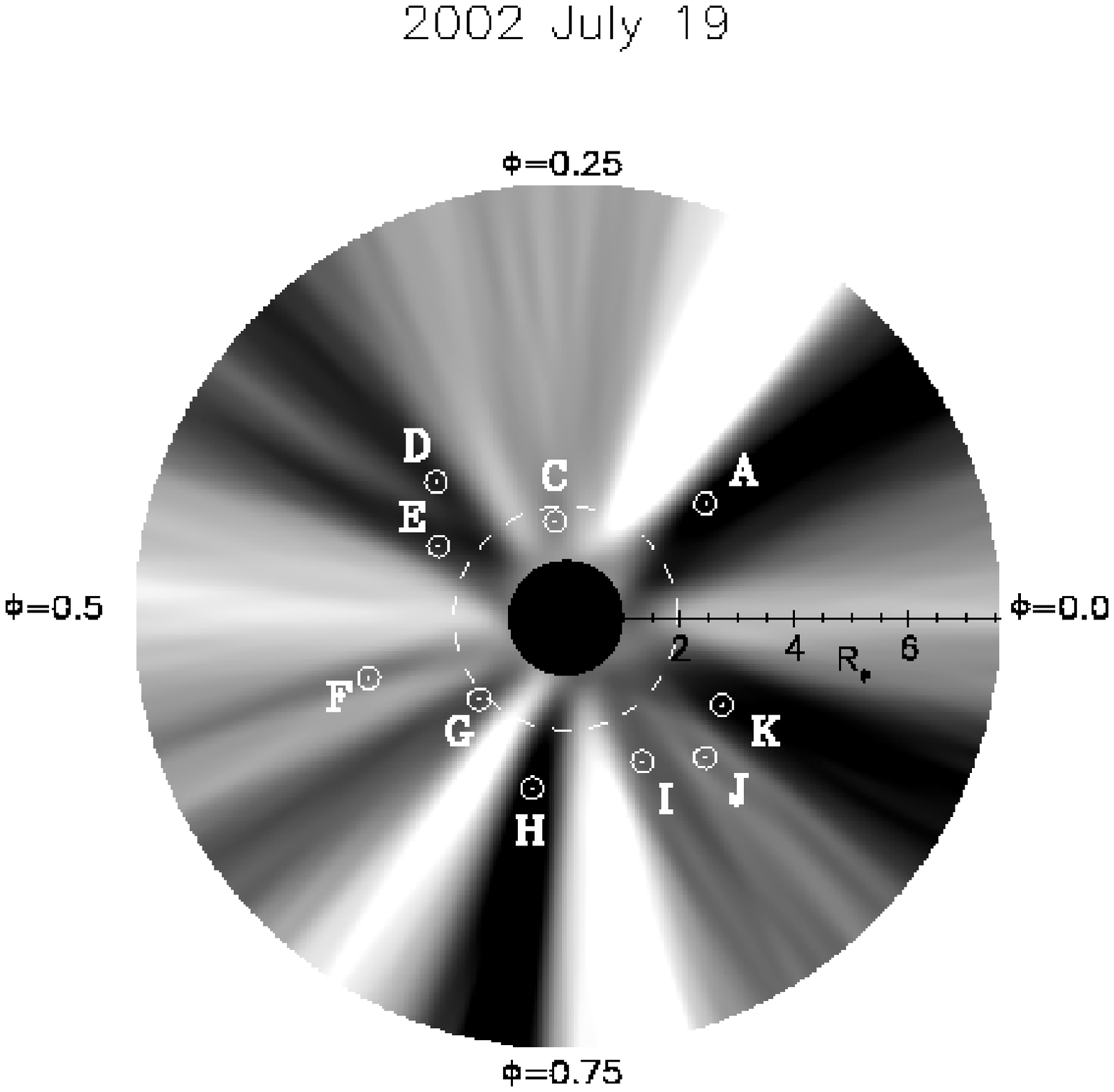} \\
    \includegraphics[width=7.0cm,angle=0]{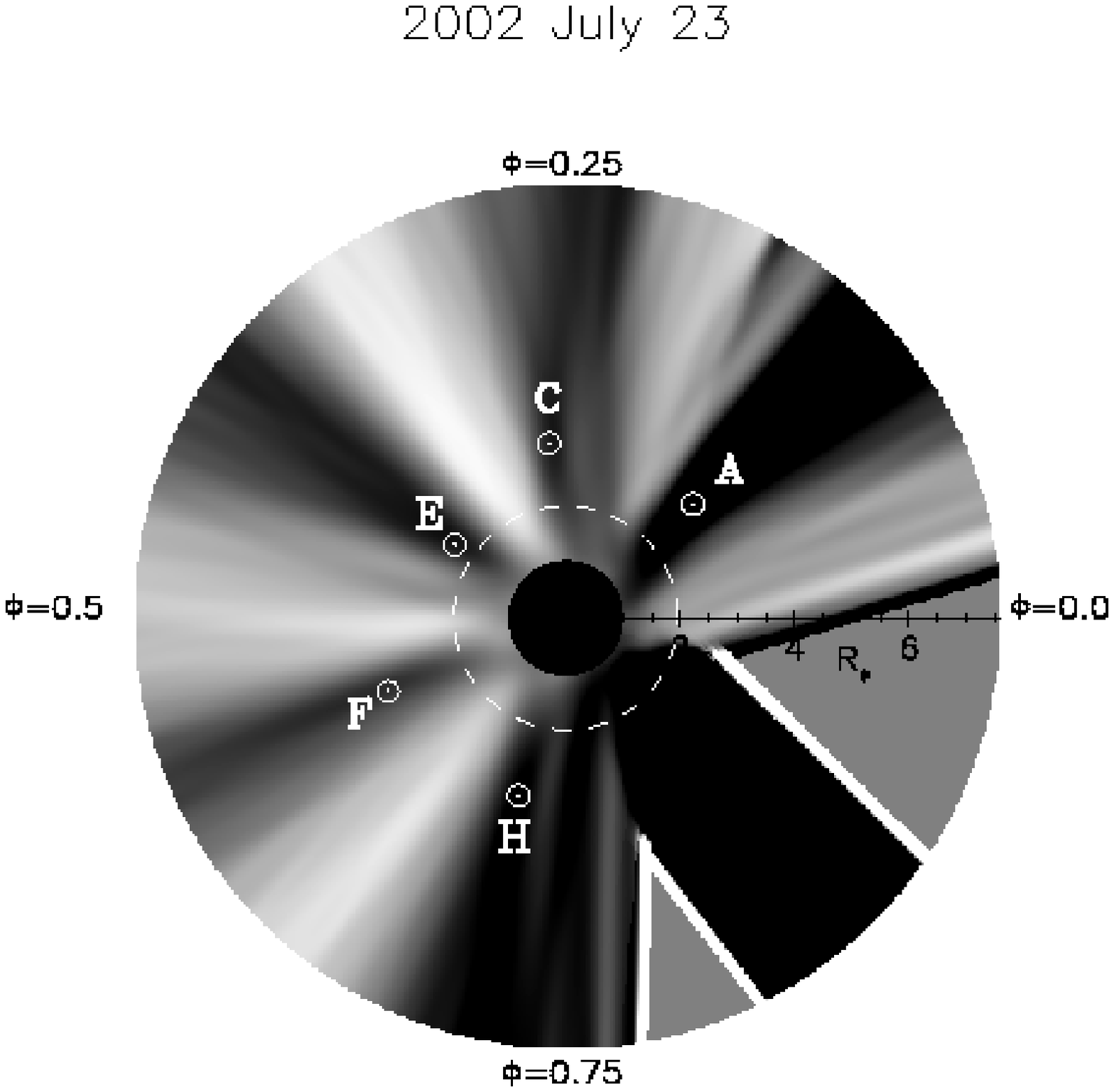} &
    \includegraphics[width=7.0cm,angle=0]{Image/njd_speedy_fig12.ps} \\
  \end{tabular}

\end{center}
\caption[Speedy Mic back projections]{Circular back projections of prominence system.  Phase increases anti-clockwise, radial distances interior to the star have been suppressed and the dashed circle shows the location of the co-rotation radius.  The grey scale represents the scale factor of the fitted Gaussian with black at -0.5 and white at +0.5. Black (negative) features show prominence absorption transients.  Circular markers show the locations of the prominences from the \chisq analysis.  The bottom right circle is a composite diagram of all prominences, those features identified on both July  18 and 19 have been joined by an arrow and crosses are July 23 data points.  The dashed lines are radial and pass through the July 18 data point to aid the eye.}
\protect\label{fig:back}
\end{figure*}

\begin{table*}
\caption{The results of the prominence tracking analysis displayed with phases of meridian crossing and calculated heights for each day of observation.  The quoted phases have a measuring uncertainty not greater than $\Delta\phi=0.002$}
\protect\label{tab:results}
\begin{center}
\begin{tabular}{ccccccccccccc}
\hline
\multicolumn{1}{c}{Prominence}	& \multicolumn{3}{c}{\it{July 18}} & \multicolumn{3}{c}{\it{July 19}}	& \multicolumn{3}{c}{\it{July 23}} 	\\
& Cycle & Phase &  $\frac{{\varpi}}{R_*}$    & Cycle &	Phase & $\frac{{\varpi}}{R_*}$  & Cycle & Phase & $\frac{{\varpi}}{R_*}$   \\

\hline	
A		& 2 & 0.102	& $2.79\pm0.11$	& 4 & 0.109 & $3.18\pm0.16$	& 15 & 0.116 & $3.00\pm0.19$\\
B		& 1 & 0.219	& $3.16\pm0.15$	&       & 		&	&& \\
C		&  &       & 		        & 4 & 0.267 & $1.71\pm0.09$	& 14 & 0.273 & $3.09\pm0.17$\\
D		& 1 & 0.343	& $2.88\pm0.14$	& 4 & 0.371 & $3.31\pm0.26$	& & &  \\
E		& 1 & 0.430	& $2.40\pm0.24$	& 4 & 0.434 & $3.00\pm0.42$	& 14 & 0.407 & $2.35\pm0.11$\\
F		& 1 & 0.535	& $3.28\pm0.18$	& 4 & 0.547 & $3.64\pm0.70$	& 14 & 0.562 & $3.39\pm0.38$\\
G		& 1 & 0.609	& $3.06\pm0.34$	& 4 & 0.620 & $2.10\pm0.12$	& & &  \\
H		& 1 & 0.703	& $2.82\pm0.37$	& 4 & 0.726 & $2.49\pm0.10$	& 14 & 0.709 & $3.24\pm0.17$\\
I		& 1 & 0.813	& $1.43\pm0.05$	& 3 & 0.828 & $2.87\pm0.40$	& & &  \\
J		& 1 & 0.871     & $3.21\pm0.22$	& 3 & 0.875 & $3.48\pm0.63$	& & &  \\
K		& & &  			        & 3 & 0.919 & $3.16\pm0.21$	& & &  \\
L		& 1 & 0.969	& $2.32\pm0.12$	& & &  			& & &  \\
\hline
\end{tabular}
\end{center}
\end{table*}

\section{Properties of the prominence system}
\protect\label{sect:proman}

The resulting prominence system is illustrated by the circular back projections displayed in Fig. \ref{fig:back}.  A total of ten prominences are found on the observable hemisphere of Speedy Mic on both nights with full phase coverage with a further five on the July 23 due to only partial phase coverage.   For reference the most studied stellar prominence system, on AB Dor, is typically quoted as having 6 to 8 prominences at any one time {(e.g. \citealt{cam1999}).   It should be noted however that obtaining full and {\it{continuous}} phase coverage of a single rotation combined with applying the improved prominence tracking technique may affect this comparison. 

In total 25 prominences are found in the range $1.4<R_c<3.6$ stellar radii above the surface.  This is the largest number of prominences observed at any single epoch on any star thus far.  Such a large number allows us to perform a statistical analysis of the prominence locations.  The distribution of prominence heights is shown in the form of a histogram in Fig. \ref{fig:hist}.

\subsection{Prominence heights and the stellar co-rotation radius}

System parameters for Speedy Mic have been constantly improved over the last decade.  Using the same dataset as this paper \cite{barnes05} were able to further constrain the period, stellar \vsini\ and axial inclination (see Table \ref{tab:sysparam}).  Such accurate parameters help us to identify Speedy Mic's co-rotation radius using:
$$R_k=\sqrt[3]{\frac{GM_*P^2}{4\pi^2}}$$
The co-rotation radius is essentially only as uncertain as the mass we adopt and even then is only proportional to the cube root of the stellar mass.   This gives the absolute position of the co-rotation radius however we need the co-rotation radius in terms of the Speedy Mic stellar radius for comparison with prominence height locations.  To calculate this we can use a simple dynamical argument assuming circular motion, using the observed \vsini, inclination and period.  We perform a Monte Carlo error analysis using the values and Gaussian-distributed $1\sigma$ uncertainties from Table \ref{tab:sysparam}.  { {Note that for the uncertainty in the period we adopt the photometric period of $0.380\pm0.004$ days as given by \cite{cutispoto97}. As this uncertainty is based on a much longer observing run and so is probably more realistic.}}  These parameters give a mean radius for Speedy Mic of $R=1.06\pm0.04$R$_\odot$, which as one might expect is considerably larger than the $0.73R_\odot$ \citep{gray92book} Zero Age Main Sequence (ZAMS) radius that the K3V spectral type would suggest.  Using Speedy Mic's determined radius and its associated error we can feed these back into another Monte Carlo trial to find the co-rotation radius in stellar radii.  For the mass we adopt $M_*=0.82\pm0.08M_\odot$.  This covers a range of masses including the $0.74M_\odot$ \citep{gray92book} K3V ZAMS and that considered by \cite{wolter05a} using pre-main sequence evolutionary models of \cite{siess2000} of $0.9M_\odot$.  The location of the equatorial co-rotation radius is found to be $R_k=1.95\pm0.07R_*$.

\begin{figure}
\begin{center}
\includegraphics[height=8cm,angle=270]{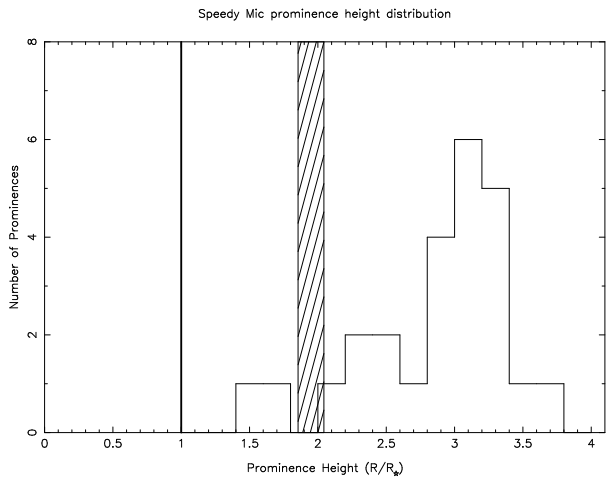}
\end{center}
\caption{Top panel: Distribution of 25 prominence heights is shown.The solid line is the stellar surface and the hashed area shows the 1$\sigma$ confidence interval on the location of the co-rotation radius.}
\protect\label{fig:hist}
\end{figure}

The histogram of prominence heights shown in Fig. \ref{fig:hist} clearly has a strong peak near $3R_*$.  This is considerably above the co-rotation radius, as illustrated in Fig. \ref{fig:hist} by the hashed region.  The actual mean of the prominence distribution is at $2.85\pm0.54R_*$ with a median of $3.00R_*$ which is less affected by the two prominences found below co-rotation.  After the $3R_*$ peak the number of prominences falls off steeply with only two prominences found above $3.4R_*$.

\subsection{Evolution of prominence system}

This dataset provides us with the opportunity to study the evolution of the prominence system on two different timescales, that of consecutive nights and then four nights later.  In Table \ref{tab:results} we attempt to match prominences found on different nights based on the simple criterion of a phase difference less than $\Delta\phi=0.03$.  { {This corresponds to approximately the extent in phase (the size) of the largest prominences.}} It is important, especially when considering consecutive nights, to note what actual physical rotational cycle each part of the phased timeseries belongs to.  The interval varies between 2 and 3 rotations between the July 18 and July 19 and this is given in Table \ref{tab:results} for each identified feature.

Using the above criterion to match the prominences we find that 8 of the 10 prominences observed on the July 18 have matching features on July 19.  Furthermore that all 8 are observed at slightly later phases on the second night than on the first night.  We also find that 6 of those 8 are observed to have moved outwards (increased in height) away from the star.  These observations are shown in graphical form in the bottom-right panel of Fig. \ref{fig:back}.

It is worth briefly mentioning individual prominences in slightly more depth.  Some prominences are easily identified on multiple nights. The strong absorption transient labelled as prominence A in Figs. \ref{fig:utseries} \& \ref{fig:back} and Table \ref{tab:results} at phase $\phi\simeq0.11$ is present at all three epochs.  Prominence F at $\phi\simeq0.54$ is present at all three epochs also.  In addition the faint absorption feature at $\phi\simeq0.87$ is clearly visible on July 18 and 19.  

Prominence B at $\phi\simeq0.219$ on July 18 seems to have disappeared by the next night.  Prominence G at $\phi\simeq0.61$ is observed at a much lower height on July 19 than on July 18.  We think this to be a newly formed prominence and in fact the feature observed on July 18 is still present but is much weaker than the new feature and is too blended in phase for the tracking technique to resolve the features unique \chisq\ minimum.

Prominence H at { {$\phi\simeq0.70$}} is faint on July 18 and at the start of observing on July 19 at phase { {$\phi\simeq0.73$}} a corresponding faint feature can just be seen finishing its track across the \ha\ profile.  At the end of that night however a much stronger absorption feature is seen at a slightly later phase.  We think this is most probably a new feature.  This is interesting as it suggests a very rapid formation timescale.  In less than 9 hours a new prominence has formed with a height of $R=2.49R_*$.

Prominence I at $\phi\simeq0.82$ is seen to experience a large increase in height between July 18 and July 19.  This particular prominence signature is open to some interpretation. The fit on July 18 is slightly ambiguous and the lower height is only marginally favoured over that of a prominence at a considerably greater height.  Then the matching of prominence I with the feature on July 19 may just be co-coincidence. 

The observations on July 23 give us a valuable opportunity to study evolution on a longer timescale, with a gap greater than 10 stellar rotations.  Somewhat surprisingly, all five  features found on July 23 have counterparts on July 19.  The heights of the features appear fairly randomly distributed (and often in between) with respect to their July 18 and 19 counterparts.  We therefore suggest that all five prominences have re-formed in the intervening time.

\section{Discussion}
\protect\label{sect:disc}

It is clear that the Speedy Mic prominence system is densely packed and relatively evenly spaced in longitude around the surface of the star.  This high level of prominence activity is what might be expected on this luminous X-ray ultra-rapid rotator.  Interestingly we find that the prominences are concentrated at $\simeq3R_*$, well above the surface of the star and at twice the height of the co-rotation radius above the stellar surface.  The traditional model of stellar prominence systems by \cite{cam1989a} describes prominences at a range of heights but with a concentration at the co-rotation radius.  This dataset certainly challenges this as 23 of the 25 prominences are found above the co-rotation radius with the peak of the prominence distribution at twice the height of the co-rotation radius above the stellar surface.  This immediately requires large extended magnetic loops that go well beyond the co-rotation radius to support such structures.

At first glance this would appear to be at odds with the high coronal densities measured with FUSE, Chandra and XMM-Newton of a range of  rapid rotators (\citealt{dupree93}; \citealt{schrijver95}; \citealt{brickhouse98}; \citealt{audard01}; \citealt{mewe01}; \citealt{young01}; \citealt{gudel01XMM}; \citealt*{sanzforcada03}; \citealt*{sanzforcada03abdor}). These measurements suggest that the coronae of rapid  rotators must be compact, since such high densities could not be  confined at large distances above the stellar surface. The lack of  significant rotation modulation in the X-ray emission can be explained  if most of the emission comes from high-latitude regions that never  pass behind the star as it rotates. This suggestion was given very  striking support by the BeppoSax observations of two flares on AB Dor that showed no rotational modulation of the emission, despite the fact  that the flares lasted for more than one rotation period of the star  \citep{maggio2000}. Modelling of the decay phase of the flare suggested  that the flaring regions were compact, and therefore must have been  located close to the rotation pole. Emission from high latitudes on  rapid rotators is not perhaps surprising given the very common  appearance of high-latitude (even polar) spots on these stars \citep{strassmeier96table}. The relationship between the high-latitude flux and the coronal distribution of X-ray emission has been determined  by  extrapolating the surface magnetic field on AB Dor obtained from  Zeeman-Doppler imaging and calculating the structure of the X-ray  emission, assuming a simple heating law (\citealt*{jardine02structure}; \citealt{jardine02xray}). The resulting X-ray emission  is confined fairly close to the stellar surface and has a large  component at high latitudes. By comparing the results of this technique  with Chandra observations obtained simultaneously with the optical  spectra used for Zeeman-Doppler imaging, \cite{hussain_chandra1_05}  showed that not only the magnitude and rotational modulation of the  X-ray emission, but also the line shifts were consistent with the  results of modelling based on field extrapolation.

If the coronae of fairly rapid rotators such as AB Dor are compact, then it might be expected that at even higher rotation rates such as that for Speedy Mic the coronal extent might be even less. For the  supersaturated star VXR45 \cite{marino03} found a large X-ray rotational modulation, showing that there is significant structure in  the corona with clearly-defined bright and dark regions. This is what  would be expected if the corona had been centrifugally stripped as a  result of the rapid rotation \citep{jardine99stripping,  jardine04stripping}, leaving a corona that is very compact with only a patchy coverage of X-ray bright regions. In the case of Speedy Mic  which appears to lie in the saturated part of the $L_x$ - period  relation it also appears unlikely that gas at coronal temperatures  could be confined by the star's magnetic field at the heights at which  the prominences are observed. The large number of prominences shows  that the confining magnetic field must still possess a significant  degree of complexity even out to these large heights.

One recent theoretical model by \cite{jardine05} puts forward a new model for prominence formation that does {\it not} require a complex global magnetic field with large extended loops in the X-ray corona supporting these prominences.  Instead they propose that the prominences are supported by the re-connection of oppositely directed wind-bearing field lines over coronal helmet streamers.  This would result in long thin highly curved loops extending out to large radii.  These could then fill up with gas that was originally part of the wind and be dragged outward in radius.  \cite{jardine05} find a maximum height for the summits of the coolest magnetic loops ($y_m$) in terms of the stellar co-rotation radius ($R_k$) given by:
$$\frac{y_m}{R_*}=\frac{1}{2}\left(-1+\sqrt{\left(1+8\left({\frac{R_k}{R_*}}\right)^3\right)}\right)$$

Using the Speedy Mic co-rotation radius $R_k=1.94R_*$ we find that $y_m=3.35R_*$.  This would seem to be consistent with the observed prominence heights.  However it should be noted that while \cite{jardine05} illustrate that such magnetic loops can achieve equilibrium they do not attempt to model the prominence structures themselves.  

The fall-off in the number of prominences detected beyond $R\simeq3R_*$ may suggest quite a different scenario to the one outlined above.  The traditional model of prominence support requires large magnetic loops in the X-ray corona.  The strongest component though at large distances away from the star is the simple dipole, however, this could only provide effective support for prominences near the equatorial plane.  Now if we assume that the observed prominences are actually in the equatorial plane then the highest prominence that would still transit the stellar disc assuming the inclination of $i=70\pm5^o$ { {is at $R\simeq3.73R_*$.  The highest detected prominence was feature F at $R=3.64R_*$ on July 19, so for Speedy Mic we cannot exclude the possibility that prominences are being supported in this manner.}}  If this were the case, the highest prominences would only start to graze the stellar disc at high latitudes resulting in a smaller absorption signature.  

\subsection{Evolution of the prominence system}

As illustrated in \S \ref{sect:proman} we have been able to identify prominences at similar longitudes on multiple epochs.  On the timescale of consecutive nights (2 to 3 rotations) we have witnessed the evolution, apparent formation and disappearance of several prominences.  The fact that prominences are observed at similar phases more than 10 rotations later is also interesting as it suggests considerable stability of the prominence supporting magnetic structures.  

Examination of the reported prominences heights and phases in Table \ref{tab:results} reveals a prominence system that appears to increase in height and lag behind in phase on the timescale of consecutive days.  It should be noted though that the apparent changes in height of many of the prominences is still consistent, within the uncertainties, with a static prominence system.  

The observed phase lags are probably more accurate measurements.  This is especially true of the narrow absorption features where the ridgeline of the absorption transients are distinct.  The problem with the larger and often stronger absorption features is that they can be comprised of multiple components.  This makes the confident re-identification of features on subsequent nights more difficult.  

The observed phase lag between the epochs could be the result of a number of factors.  \cite{donati97ab} and \cite{donati99ab} attribute similar observations on AB Dor to the differential rotation of the prominence footprints.  This then allows the latitude of the footprints to be located.  In doing so though they assume that the supporting magnetic field lines are rigid and thus maintain prominence co-rotation.  For Speedy Mic however we find that the phase lags between the consecutive days are generally of the order of $\Delta\phi\simeq0.01$.  This results in prominence periods that are longer than the surface rotation rate at any latitude on the stellar surface and so cannot be explained by differential rotation.  This may suggest that the stellar magnetic field is unable to enforce co-rotation.  If we consider this in the context of prominences moving away from the star then the observed phase lag between consecutive days could be the result of the prominences being dragged back in phase due to conservation of angular momentum.  As such the phase lags observed between consecutive days have the potential to inform us about the rigidity of the confining magnetic field.

If the above model is correct then we need to explain why there is more noticeable phase lags between epochs on Speedy Mic than on AB Dor.  This may simply be due to the location of the prominences relative to the co-rotation radius.  We find that prominences group around a height which is twice that of the co-rotation radius above the stellar surface yet, as already discussed, prominences on AB Dor are found at or just above the co-rotation radius.  Therefore it maybe more difficult for the magnetic field to enforce co-rotation of the prominence structures on Speedy Mic.  Larger phase lags have however been observed before on very high prominences on AB Dor.  During observations in 1994 \cite{cam1999} found that a prominence they observed moving radially outward from 3 or 4 stellar radii to 8 stellar radii lagged behind in phase.  The authors attributed this to an attempt by the prominence to conserve angular momentum.

Examining now the two prominences confidently identified on all three epochs, prominences A and F, we find that they both have relatively large phase lags between July 18 and 19 (two and three stellar rotations respectively).  Then approximately ten rotations later we observe only a similar additional lag in phase on July 23.  This would suggest that the prominences have reformed at least once in the intervening ten rotations.  This slower lag in phase may indeed be due to differential rotation of the prominence footprints. { { \cite{barnes05} fitted a solar like differential rotation law to starspots observed on \speedy\ of the form  \mbox{$\Omega(\theta) = \Omega_0 - \Delta\Omega\rm{sin}^2(\theta)$}, (where $\theta$ is the stellar latitude and $\Omega_0$ the equatorial rotation velocity).  The values obtained were \hbox{\mbox{$\Omega_0 = 16.5361$}} \hbox{\mbox{$\pm 0.0006 $} \radday}\ and \hbox{\mbox{$\Delta\Omega = 0.033 \pm 0.003$ }}.  We can use this to attempt to locate the latitudes of the prominence footprints, as was done for AB Dor by \cite{donati97ab}.  The observed phase lag over the ten rotations corresponds to a period for prominence A that would place it at a latitude of $\simeq50$\degs\ on the stellar surface.  Similarly prominence F would correspond to a latitude of $\simeq80$\degs. The associated uncertainties on these latitudes come from the uncertainty in the recurrence time and the uncertainty in the strength of the differential rotation.  Resulting in an uncertainty of $\simeq10$\degs for prominence A and $\simeq20$\degs for prominence F. These are similar latitudes to those found by \cite{donati97ab}, suggesting that the prominence footprints are located well above the stellar equator.}}

Given the picture of prominence evolution that we have just described it seems that much care must be exercised in performing such an analysis of the prominence footprint locations.  If the phase at which a prominence is observed is a function of its height then this must be taken into account.  In order for a valid assessment of the phase lag due to differential rotation we need to observe the prominence at the same height on multiple epochs.  It seems that this would require observing a prominence that has reformed at a similar phase and is at a similar stage in its height evolution.  Due to our lack of understanding of the prominence formation mechanism we cannot be sure that we are really observing the same loop structures.  This is brought home by the data in Table \ref{tab:results} where we find that the two prominences (G and H) thought to have reformed between July 18 and 19 were actually observed at slightly later phases than the original structures.  This may point to a re-organisation of field at the prominence footprints such that the newly formed loops are not at exactly the same locations as the previous ones.  If so then this would invalidate our analysis of the prominence footprint latitudes.

\section{Conclusions}
\protect\label{sect:conc}

The improved prominence tracking technique allows for the fast, relatively automated detection and tracking of multiple prominences as they cross the stellar \vsini\ profile.  { {This involved a matched filter analysis technique, fitting a gaussian model with a simple stellar limb-weighting function by a \chisq\ analysis.  The Speedy Mic \ha\ timeseries has revealed one of the most (if not the most) densely packed and active prominence systems seen to date.  }}

We have shown that the distribution of prominence heights peaks considerably above the co-rotation radius.  This provides further evidence against the traditional model of prominence support via large magnetic loops in the X-ray corona.  Having discussed an alternative theoretical model of prominence support it would be very interesting to carry out Zeeman Doppler Imaging of Speedy Mic to map the surface magnetic field. { {In principle this could be done using SEMPOL at the AAT, but Speedy Mic is too faint to avoid excessive phase smearing.}} Unfortunately Speedy Mic has too southerly a declination for observation with the new echelle spectropolarimeter `ESPaDOnS' (see \citealt{donati2003EsP}) on the 4-m Canada-France-Hawaii Telescope. Ideally, an analagous high throughput spectropolarimeter at a 4-m or 8-m Southern hemisphere telescope is required. 

This study provides tentative evidence for a coherent picture of prominence evolution on \speedy.  Whether prominences are actually moving away from the star and lagging behind in phase as they attempt to conserve their angular momentum is still subject to interpretation.  { {In order to more confidently distinguish between this effect and that of the differential rotation of the prominence footprints further observations will be necessary.  These should focus on tracking the evolution of individual prominence structures with the shortest time intervals possible.  If we can obtain repeat phase coverage on three or more consecutive nights then we could hope to track multiple prominences through their entire evolution. Thus we could hope to define the relationship between prominence height and observed phase lag.  This should give us an insight into the strength and nature of the magnetic support and further constrain angular momentum loss through prominence ejection.}}

While most of this paper has focussed on the rapid absorption transients interpreted as stellar prominences there are clearly other interesting things happening in the \ha\ line.  The raw timeseries in Fig. \ref{fig:rtseries} shows slower variations in the underlying \ha\ absorption.  These seem to split the star into two longitude bands, for one half of the observed stellar rotation we see dark regions of absorption while for the half we see much less absorption.  This is quite different from the \ha\ profile of AB Dor, where no such obvious contrast has been observed and as such requires further study.

\section{ACKNOWLEDGEMENTS}

The data in this paper were reduced using {\sc starlink} software packages.  NJD acknowledges the financial support of a UK PPARC studentship.  We would like to thank our referee U. Wolter for his helpful comments on improving the clarity of the paper.

\bibliographystyle{mn2e}
\bibliography{iau_journals,master,ownrefs,njd2}

\end{document}